\documentclass[showpacs,reprint]{revtex4}
\usepackage{amsfonts}
\usepackage{amssymb}
\usepackage{amsmath}
\usepackage{hyperref}
\usepackage{latexsym}
\usepackage[dvips]{graphicx}
\usepackage{epsf}
\usepackage{multirow}

\begin{document}

\title{On the possibility of the generation of an artificial wormhole}
\author{ E.P. Savelova}\email{sep$\_$ 22.12.79@inbox.ru} \author{A.A. Kirillov}
 \affiliation{Dubna International
University of Nature, Society and Man, Universitetskaya Str. 19,
Dubna, 141980, Russia }
\date{}

\begin{abstract}
We model spacetime foam by a gas of virtual wormholes. Then
applying an external field one may change the density of virtual
wormholes and try to organize a wormhole-like structure in space.
The relation between an additional distribution of virtual
wormholes and the external field is considered for the homogeneous
case. We show that the external fields suppress the density of
virtual wormholes which sets an obstruction for creating an actual
wormhole in a straightforward fashion. We also present a rough
idea of a more complicated model for the artificial creation of a
wormhole.
\end{abstract}

\maketitle

\section{Introduction}

It is well known that physics of wormholes involves the two important
features. The first one is that wormholes violate the null energy condition
(NEC) \cite{VisX} which requires the presence of an exotic (which does not
exist in lab experiments) matter. The second important feature is the fact
that topology changes are rigorously forbidden in classical general
relativity, e.g., see the theorem of Geroch in Ref. \cite{ger}.
Nevertheless, there exists a widely spread misunderstanding that it is
enough to prepare an exotic NEC - violating matter and this may allow to
create a wormhole artificially. This is surely not true, for it does not
matter which sort of the stress energy tensor we may use as a source in
classical general relativity, topology will remain the same. An exotic
energy produces only an exotic dynamics but not any topology change. In
particular, the possibility to create a new universe (or baby universe) by a
man-made process discussed first in Refs. \cite{bubbl} \ and recently \ in
connection to NEC - violating theories in Ref. \cite{R13} does not assume
any topology change but rather an inflating dynamics for a small portion of
space.

The first rigorous model which may allow to describe the creation of an
artificial wormhole was suggested recently in Ref. \cite{KS13a}. This model
is based on quantum gravity effects, namely, on the spacetime foam picture
\cite{wheelerX,H78X}. It is assumed that at very small (Planckian) scales
spacetime is filled with a gas of virtual wormholes. The virtual wormhole
represents a quantum topology fluctuation (tunnelling between different
topologies) which takes place at very small (Planckian) scales and lasts for
a very short (Planckian) period of time. It does not obey to the Einstein
equations and, therefore, violates readily NEC. Virtual wormholes are
described by Euclidean wormhole configurations which were first suggested in
Refs. \cite{Loss1}. Our model is based on the fact that a coherent set of
virtual wormholes may work as an actual wormhole \cite{S13,S14} (in somewhat
different context analogous idea was discussed in Ref. \cite{Knot}). Thus,
by applying an external classical field one may govern the intensity of such
topology fluctuations (i.e., the density of virtual wormholes) and try to
organize an artificial wormhole. In this case a topology change does not
occur at all and this is not in the contradiction with the Geroch theorem.

We in Ref. \cite{KS13a} missed however the sign of the Green function and
assumed the delta-correlation for the bias function (i.e., for the
additional sources generated on throats of wormholes) which, in general, is
not correct. In the present paper we consider a more general but still a
homogeneous case. We show how an external field relates to topology
fluctuations (i.e., to the perturbations in the number density of virtual
wormholes). The naive expectations, e.g., that a homogeneous external
electric field may redirect virtual wormholes along the direction of the
electric lines, seem to be not working. Indeed, if it were the case, such a
phenomenon would be observed long ago in laboratory experiments. Instead, it
turns out that external fields somewhat suppress the number density of
wormholes. Such a behaviour becomes clear if we recall that virtual
wormholes diminish the energy density of zero-point fluctuations (i.e., the
vacuum energy density) \cite{KS10X}. The external classical field carries
always the positive energy density and, therefore, in the presence of any
external classical field the vacuum energy density becomes somewhat higher.
Which means that the number density of virtual wormholes should be somewhat
smaller.

At first glance the fact that an external classical field suppresses the
density of virtual wormholes means that the straightforward creation of an
actual wormhole meets some obstruction (which seems to agree with Ref. \cite%
{R13} where an obstruction for creating a universe in the laboratory was
reported). However it is clearly not the case. One may consider a more
complicated models, e.g., in the case of a metastable vacuum state (which
probably took place in the very early Universe) or (more realistically) one
may use a dual (in some sense) field which suppresses a some number density
of wormholes save a coherent set\footnote{%
We recall that virtual wormholes diminish the vacuum energy density \cite%
{KS10X} and make it finite and, therefore, it is merely impossible to
suppress all virtual wormholes; for it would require an infinite energy. }.
We see that such a dual fields involve  more energy density and have the
more complicated configurations. This may explain why such a phenomenon
(creating of a wormhole-like structure) was not observed so far. We should
also reserve the possibility that inhomogeneous external fields (which are
still out of our consideration) may produce a consistent model for creating
an artificial wormhole which requires the further study.

The negative perturbation in the number density of wormholes in the presence
of external fields may however lead to a positive shift in the value of the
cosmological constant \cite{KS10X} (we assume that in the absence of
external fields the cosmological constant should be exactly zero). Whether
it is possible to use such a phenomenon to explain the observed acceleration
of the Universe (or to implement the ideas of Ref. \cite{R13}) or not
requires the further study.

\section{Green function in a gas of virtual wormholes}

Consider now the simplest massless scalar field and construct the Green
function in the presence of a gas of wormholes. The Green function obeys the
Laplace equation
\begin{equation*}
-\Delta G\left( x,x^{\prime }\right) =\delta \left( x-x^{\prime }\right)
\end{equation*}%
with proper boundary conditions at throats (we require $G$ and $\partial
G/\partial n$ to be continual at throats). The Green function for the
Euclidean space is merely
\begin{equation*}
G_{0}(x,x^{\prime })=\frac{1}{4\pi ^{2}\left( x-x^{\prime }\right) ^{2}}
\end{equation*}%
(and $G_{0}\left( k\right) =1/k^{2}$ for the Fourier transform). In the
presence of a single wormhole which connects two Euclidean spaces this
equation admits the exact solution e.g., see details in Ref. \cite{S12ef,S14}%
. In this case the wormhole is described by the metric ($\alpha =1,2,3,4$)%
\begin{equation}
ds^{2}=h^{2}\left( r\right) \delta _{\alpha \beta }dx^{\alpha }dx^{\beta },
\label{X_wmetr}
\end{equation}%
where
\begin{equation}
h\left( r\right) =1+\theta \left( a-r\right) \left( \frac{a^{2}}{r^{2}}%
-1\right)
\end{equation}%
and $\theta \left( x\right) $ is the step function. Such a wormhole has
vanishing throat length. Indeed, in the region $r>a$, $h=1$ and the metric
is flat, while the region $r<a$, with the obvious transformation $y^{\alpha
}=\frac{a^{2}}{r^{2}}x^{\alpha }$, is also flat for $y>a$. Therefore, the
regions $r>a$ and $r<a$ represent two Euclidean spaces glued at the surface
of a sphere $S^{3}$ with the center at the origin $r=0$ and radius $r=a$. \

For the outer region of the throat $S^{3}$ the source $\delta \left(
x-x^{\prime }\right) $ generates a set of multipoles placed in the center of
sphere which gives the corrections to the Green function $G_{0}$ in the form
(we suppose the center of the sphere at the origin)%
\begin{equation}
\delta G=-\frac{1}{4\pi ^{2}x^{2}}\sum_{n=1}^{\infty }\frac{1}{n+1}\left(
\frac{a}{x^{\prime }}\right) ^{2n}\left( \frac{x^{\prime }}{x}\right)
^{n-1}Q_{n},
\end{equation}%
where $Q_{n}=\frac{4\pi ^{2}}{2n}\sum_{l=0}^{n-1}\sum_{m=-l}^{l}Q_{nlm}^{%
\ast \prime }Q_{nlm}$ and $Q_{nlm}\left( \Omega \right) $ are
four-dimensional spherical harmonics e.g., see \cite{fockX}. And analogous
expression exists for the inner region of the sphere $S^{3}$  \cite%
{S12ef,S14}. In the present paper we shall consider a dilute gas
approximation and, therefore, it is sufficient to retain the lowest
(monopole) term only. We point out that the monopole terms coincide for the
inner and outer regions of the throat $S^{3}$. A single wormhole which
connects two regions in the same space is a couple of conjugated spheres $%
S_{\pm }^{3}$ of the radius $a$ with a distance $\vec{X}=\vec{R}_{+}-\vec{R}%
_{-}$ between centers of spheres. So the parameters of the wormhole are $\xi
=(a,R_{+},R_{-})$. The interior of the spheres is removed and surfaces are
glued together. Then the proper boundary conditions (the actual topology)
can be accounted for by adding the bias of the source
\begin{equation}
\delta (x-x^{\prime })\rightarrow N\left( x,x^{\prime }\right) =\delta
(x-x^{\prime })~+b\left( x,x^{\prime }\right) .
\end{equation}%
In the approximation $a/X\ll 1$ (e.g., see also \cite{KS07X}) the bias takes
the form
\begin{equation}
b_{1}\left( x,x^{\prime },\xi \right) =\frac{a^{2}}{2}\left( \frac{1}{\left(
R_{-}-x^{\prime }\right) ^{2}}-\frac{1}{\left( R_{+}-x^{\prime }\right) ^{2}}%
\right) \times   \label{X_b1}
\end{equation}%
\begin{equation*}
\times \left[ \delta ^{4}(x-R_{+})-\delta ^{4}(x-R_{-})\right]
\end{equation*}%
We expect that virtual wormholes have throats $a\sim \ell _{pl}$ of the
Planckian size, while in the present paper we are interested in much larger
scales. Therefore, the form (\ref{X_b1}) is sufficient for our aims. However
this form is not acceptable in considering the short-wave behavior and
vacuum polarization effects (e.g., the stress energy tensor). In the last
case one should account for the finite value of the throat size and replace
in (\ref{X_b1}) the point-like source with the surface density (induced on
the throat) e.g., see for details \cite{KS10X}, $\delta ^{4}(x-R_{\pm
})\rightarrow \frac{1}{2\pi ^{2}a^{3}}\delta (\left\vert x-R_{\pm
}\right\vert -a).$

In the rarefied gas approximation the bias function for the gas of wormholes
is additive, i.e.,
\begin{equation}
b\left( x,x^{\prime }\right) =\sum b_{1}\left( x,x^{\prime },\xi _{i}\right)
=\int b_{1}(x,x^{\prime },\xi )F(\xi )d\xi ,  \label{X_b2}
\end{equation}%
where $F\left( \xi \right) $ is the number density of wormholes in the
configuration space which is given by%
\begin{equation}
F\left( \xi \right) =\sum\limits_{i=1}^{N}\delta \left( \xi -\xi _{i}\right)
.  \label{X_F}
\end{equation}

In the vacuum case this function has a homogeneous distribution $\rho (\xi )=
$ $<0|F\left( \xi \right) |0>=\rho \left( a,X\right) $, then for the mean
bias $\overline{b}=<0|b|0>$ \ we find
\begin{equation}
\overline{b}\left( x-x^{\prime }\right) =\int a^{2}\left( \frac{1}{R_{-}^{2}}%
-\frac{1}{R_{+}^{2}}\right) \delta ^{4}(x-x^{\prime }-R_{+})\rho \left(
a,X\right) d\xi   \label{X_bx}
\end{equation}%
Consider the Fourier transform $\rho \left( a,X\right) =\int \rho \left(
a,k\right) e^{-ikX}\frac{d^{4}k}{\left( 2\pi \right) ^{4}}$ then we find for
$b\left( k\right) =\int b\left( x\right) e^{ikx}d^{4}x$ the expression
\begin{equation}
\overline{b}\left( k\right) =\frac{4\pi ^{2}}{k^{2}}\int a^{2}\left( \rho
\left( a,k\right) -\rho \left( a,0\right) \right) da,  \label{X_bk}
\end{equation}%
which forms the background cutoff function $\overline{N}\left( k\right) =1+%
\overline{b}\left( k\right) $, so that the regularized vacuum Green function
$G_{reg}\left( k\right) $ has the form%
\begin{equation}
G_{reg}\left( k\right) =\frac{1}{k^{2}}\overline{N}\left( k\right) .
\label{X_GF}
\end{equation}%
General properties of the cutoff is that $\overline{N}\left( k\right)
\rightarrow 0$ as $k\gg k_{pl}$ and $\overline{N}\left( k\right) \rightarrow
const\leq 1$ in the low energy limit as \thinspace $k\ll k_{pl}$. The next
our aim is to find the change of the bias in the presence of an external
field. To this end we consider the structure of the partition function.

\section{Structure of the generating functional}

To relate the additional distribution of virtual wormholes and an external
current we consider now the generating functional (the partition function)
which is used to generate all possible correlation functions in quantum
field theory (and the perturbation scheme when we include interactions)
\begin{equation}
Z_{total}\left( J\right) =\sum\limits_{\tau }\sum\limits_{\varphi }e^{-S_{E}}
\end{equation}%
where the sum is taken over field configurations $\varphi $ and topologies $%
\tau $ (wormholes), the Euclidean action is
\begin{equation}
S_{E}=-\frac{1}{2}\left( \varphi \Delta \varphi \right) -\left( J\varphi
\right) ,  \label{X_act}
\end{equation}%
and we use the notions $\left( J\varphi \right) =\int J\left( x\right)
\varphi \left( x\right) d^{4}x$. Here $J$ denotes an external current. The
sum over field configurations $\varphi $ can be replaced by the integral $%
Z^{\ast }\left( J\right) =\int \left[ D\varphi \right] e^{-S_{E}}$ which
gives%
\begin{equation}
Z^{\ast }=Z_{0}(G)e^{\frac{1}{2}\left( JGJ\right) },  \label{X_gf2}
\end{equation}%
where $Z_{0}(G)=\int \left[ D\varphi \right] e^{\frac{1}{2}\left( \varphi
\Delta \varphi \right) }$ is the standard expression and $G=G\left( \xi
_{1},...,\xi _{N}\right) $ is the Green function for a fixed topology, i.e.,
for a fixed set of wormholes $\xi _{1},...,\xi _{N}$.

Consider now the sum over topologies $\tau $. To this end we restrict with
the sum over the number of wormholes and integrals over parameters of
wormholes:
\begin{equation}
\sum\limits_{\tau }\rightarrow \sum\limits_{N}\int
\prod\limits_{i=1}^{N}d\xi _{i}=\int \left[ DF\right]   \label{X_ts}
\end{equation}%
where $F$ is given by (\ref{X_F}). We point out that in general the
integration over parameters is not free (e.g., it obeys the obvious
restriction $\left\vert \vec{R}_{i}^{+}-\vec{R}_{i}^{-}\right\vert \geq
2a_{i}$). This defines the generating function as
\begin{equation}
Z_{total}\left( J\right) =\int \left[ DF\right] Z_{0}(G)e^{\frac{1}{2}\left(
JGJ\right) }.
\end{equation}%
In this expression $J$ may serve as an external current which generates an
external classical background field $\varphi _{ext}=GJ$.

In the vacuum case virtual wormholes have a homogeneous distribution.
Therefore, if we remain within homogeneous distributions $F$ (\ref{X_F})
(which sets also additional restrictions on possible external fields), then
the structure of the partition function $Z_{total}\left( J\right) $ can be
investigated a little bit further. Indeed, for homogeneous states in the
Fourier representation the bias $N(x,x^{\prime },\xi )\rightarrow
N(k,k^{\prime },\xi )$ takes the form
\begin{equation*}
N(k,k^{\prime })\ =N(k,\xi )\delta (k-k^{\prime }),
\end{equation*}%
then the true Green function takes the form (compare with (\ref{X_GF})) $%
G\left( k\right) =G_{0}\left( k\right) N(k,\xi )$ and for the total
partition function we find
\begin{equation}
Z_{total}\left( J_{ext}\right) =\int \left[ DN(k)\right] e^{-I(N(k))}e^{%
\frac{1}{2}\sum \frac{1}{k^{2}}N(k)\left\vert J_{k}\right\vert ^{2}},
\label{X_ztot}
\end{equation}%
where $\sum_{k}=\int \frac{d^{4}k}{\left( 2\pi \right) ^{4}}$ and $\left[ DN%
\right] =\prod\limits_{k}dN_{k}$ . The functional $I(N)$ comes from the
integration measure (which includes the Jacobian of transformation from $%
F\left( \xi \right) $ to $N\left( k\right) $)
\begin{equation*}
e^{-I(N)}=\int \left[ DF\right] Z_{0}(N(k,\xi ))\delta \left( N\left(
k\right) -N\left( k,\xi \right) \right)
\end{equation*}%
and has the sense of the action for the bias function $N\left( k\right) $.
In the true vacuum case $J=0$ and by means of using the expression (\ref%
{X_ztot}) we find the two-point Green function in the form
\begin{equation}
G\left( k\right) =\frac{\overline{N}(k)}{k^{2}}  \label{X_grfm}
\end{equation}%
where $\overline{N}(k)$ is the cutoff function (the mean bias) which is
given by
\begin{equation*}
\overline{N}(k)=<0|N\left( k\right) |0>=\frac{1}{Z_{total}\left( 0\right) }%
\int \left[ DN\right] e^{-I\left( N\right) }N\left( k\right) .
\end{equation*}

The action $I(N)$ can be expanded as\footnote{%
In this expansion the mean cutoff $\overline{N}(k)$ is merely the solution
of $\frac{\delta I\left( N\right) }{\delta N\left( k\right) }=0$.}
\begin{equation}
I(N)=I(\overline{N})+\frac{1}{2}\sum_{kp}M_{k,p}\Delta N\left( k\right)
\Delta N^{\ast }\left( p\right) +...  \label{X_actn}
\end{equation}%
where $\Delta N\left( k\right) =N\left( k\right) -\overline{N}(k)$ and the
kernel $M_{k,p}$ $=1/\sigma _{k,p}^{2}$ can be expressed via the dispersion
of vacuum topology fluctuations as
\begin{equation*}
\sigma _{k,p}^{2}=\overline{\Delta N\left( k\right) \Delta N^{\ast }\left(
p\right) }=\frac{1}{Z_{total}\left( 0\right) }\int \left[ DN\right]
e^{-I\left( N\right) }\Delta N\left( k\right) \Delta N^{\ast }\left(
p\right) .
\end{equation*}%

\section{Topology fluctuations in the presence of exertnal fields}

Consider now topology fluctuations in the presence of an external current.\
In the presence of an external current $J^{ext}$ the intensity of topology
fluctuations changes. Indeed using (\ref{X_ztot}), (\ref{X_actn}) we find
\begin{equation}
I(N,J^{ext})=I(\overline{N})+\frac{1}{2}\sum_{k,p}M_{k,p}\Delta N\left(
k\right) \Delta N\left( p\right) -\frac{1}{2}\sum_{k}\frac{1}{k^{2}}%
\left\vert J_{k}^{ext}\right\vert ^{2}N(k)
\end{equation}%
which gives
\begin{equation}
I(N,J^{ext})=I(\overline{N},J^{ext})+\frac{1}{2}\sum_{kp}M_{k,p}\delta
N\left( k\right) \delta N\left( p\right)
\end{equation}%
where $\delta N\left( k\right) =N\left( k\right) -\overline{N}(k,J^{ext})$,
\begin{equation}
I(\overline{N},J^{ext})=I(\overline{N})-\frac{1}{8}\sum_{k,p}\sigma
_{k,p}^{2}\left( \frac{1}{k^{2}}\left\vert J_{k}^{ext}\right\vert
^{2}\right) \left( \frac{1}{p^{2}}\left\vert J_{p}^{ext}\right\vert
^{2}\right) ,
\end{equation}%
and $\overline{N}(k,J^{ext})=\overline{N}(k)+\delta b_{k}\left( J\right) $
with
\begin{equation}
\delta b_{k}\left( J\right) =\frac{1}{2}\sum_{p}\sigma _{k,p}^{2}\frac{1}{%
p^{2}}\left\vert J_{p}^{ext}\right\vert ^{2}.  \label{b3}
\end{equation}

We recall that this expression works only in the case when the external
current does not destroy the homogenity of vacuum state but still it is
exact and contains yet unknown kernel $\sigma _{k,p}^{2}$. It is possible to
get an analogous expression by perturbation method without the restriction
to homogeneity of the vacuum state (i.e., in the presence of an arbitrary
external field)  but it has (as well as the exact form of the kernel $\sigma
_{k,p}^{2}$) too complicated structure and we present it elsewhere \cite%
{KS14}. Here we only briefly describe the way to get its exact expression.
First, one has to notice that the bias has the structure $N(x,x^{\prime
})=\delta (x-x^{\prime })+b(x,x^{\prime })$ where $b$ depends on wormhole
parameters $\xi $ and is given by ((\ref{X_b2}). Therefore, the dispersion $%
\sigma $ reduces to $\sigma _{1,2}^{2}=<0|\Delta b_{1}\Delta b_{2}|0>$,
while the former may be expressed via the two-point correlation function for
the wormhole number density in the configuration space (\ref{X_F}) as $%
\sigma _{1,2}^{2}=\int b_{1}(\xi )b_{2}(\xi ^{\prime })\rho (\xi ,\xi
^{\prime })d\xi d\xi ^{\prime }$, where $\rho (\xi ,\xi ^{\prime
})=<0|\Delta F(\xi )\Delta F(\xi ^{\prime })|0>$. In a gas of wormholes $%
\rho $ has the structure $\rho (\xi ,\xi ^{\prime })=\rho (\xi )\delta (\xi
-\xi ^{\prime })+\nu (\xi ,\xi ^{\prime })$, where $\rho (\xi )=\rho \left(
a,X\right) =<0|F(\xi )|0>$ \ is the mean vacuum density which we used in (%
\ref{X_bx}) and $\nu $ describes correlations. In the rarefied gas
approximation $\nu $ may be neglected to the leading order, while in general
the dependence on $\xi $ can be found directly from (\ref{X_gf2}) as $\rho
(\xi _{1},\xi _{2})\sim Z_{0}(G(\xi _{1},\xi _{2}))$.

\section{Conclusion}

The expression (\ref{b3}) shows that $\delta b_{k}\left( J\right) >0$
contrary to the naive expectations. In particular, for the background bias (%
\ref{X_b1}), (\ref{X_bk}) \ the value $\overline{N}(k)$ is always less than
unity and $\overline{b}\left( k\right) <0$. This means that external fields
always suppress the density of virtual wormholes. Indeed let us assume the
homogeneous perturbation in the presence of $J\neq 0$
\begin{equation}
\delta \rho \left( a,X\right) =\delta n\delta \left( a-a_{0}\right) \frac{1}{%
2}\left( \delta ^{4}\left( X-r_{0}\right) +\delta ^{4}\left( X+r_{0}\right)
\right) ,  \label{X_NF}
\end{equation}%
where $\delta n=\delta N/V$ is the change in the density of wormholes. We
point out that in the vacuum case the background density of wormholes is
always positive $n\geq 0$, while the value $\delta n$ admits both signs. The
above distribution corresponds to a set of wormholes with the throat size $%
a_{0}$, oriented along the same direction $r_{0}$ and with the distance
between throats $r_{0}=\left\vert R_{+}-R_{-}\right\vert $. Then $\delta
\rho \left( a,k\right) $ reduces to $\delta \rho \left( a,k\right) =\delta
n\delta \left( a-a_{0}\right) \cos \left( kr_{0}\right) $, where $\left(
kr_{0}\right) =k_{\mu }r_{0}^{\mu }$. Thus from (\ref{X_bk}) we find $\delta
b_{k}(J)=-\delta na_{0}^{2}\frac{4\pi ^{2}}{k^{2}}\left( 1-\cos \left(
kr_{0}\right) \right) $ and substituting this into (\ref{b3}) we may define
the external field $\varphi =GJ$ which generates such a perturbation $\delta
\rho $ in the vacuum distribution of wormholes. The property $\delta
b_{k}\left( J\right) >0$ means that the perturbation in the wormhole number
density is always negative $\delta n(J)<0$.

According to discussions in Ref. \cite{S14} there may exist a situation when
the speed of light remains isotropic (in an appropriate external field) but
slightly exceeds the vacuum value. Such a phenomenon can be in principle
observed in an experiment analogous to that by the OPERA Collaboration \cite%
{lightX}.

The nagative value $\delta n(J)<0$ means that the possibility to create a
homogeneous wormhole-like structure in space in a straightforward fashion
meets some obstructions. The way out of this difficulty is to consider a
metastable vacuum state which may appear in the presence of an inhomogeneous
external field (which require the further study). Still there is a more
simple situation when the external field suppressess the dual distribution
of wormholes, e.g., the distribution in the form
\begin{equation}
\delta \rho _{dual}\left( a,X\right) =\delta n-\delta \rho \left( a,X\right)
,
\end{equation}%
where $\delta \rho \left( a,X\right) $ is given by (\ref{X_NF}).
Then the negative value $\delta n<0$ will mean the presence of a
some excess of virtual wormholes with the distribution
(\ref{X_NF}), i.e., the formation of the wormhole-like structure
discussed in Ref. \cite{S14}. We recall that to suppress all
virtual wormholes simply impossible since it would require an
infinite energy.

We acknowledge V. 
Berezin 
for useful discussions.

\end{document}